\documentclass[doublecol]{epl2} 
\usepackage{epstopdf}
\usepackage{ulem}

\title{A nonextensive insight to the stellar initial mass function}
\shorttitle{A nonextensive insight to the stellar initial mass function} 

\author{D. B. de Freitas\inst{1}, R. T. Eufr\'asio\inst{2,3}, M. M. F. Nepomuceno\inst{4,5} \and J. R. P. da Silva\inst{5}}
\shortauthor{de Freitas et al.}

\institute{                    
	\inst{1} Departamento de F\'{\i}sica, Universidade Federal do Cear\'a, Caixa Postal 6030, Campus do Pici, 60455-900 Fortaleza, Cear\'a, Brasil\\
	\inst{2}{Department of Physics, University of Arkansas, 226 Physics Building, 825 West Dickson Street, Fayetteville, AR 72701, USA}\\
    \inst{3}{NASA Goddard Space Flight Center, Greenbelt, MD 20771, USA}\\
	\inst{4} Departamento de Ci\^encia e Tecnologia, Universidade Federal Rural do Semi-\'Arido, Campus Cara\'ubas, Rio Grande do Norte, Brasil. \\
	\inst{5} Departamento de F\'isica, Universidade do Estado do Rio Grande do Norte, Mossor\'o, Brasil. \\
}

\abstract{In the present paper, we propose that the stellar initial mass distributions as known as IMF are best fitted by $q$-Weibulls that emerge within nonextensive statistical mechanics. As a result, we show that the Salpeter's slope of $\sim$2.35 is replaced when a $q$-Weibull distribution is used. Our results point out that the nonextensive entropic index $q$ represents a new approach for understanding the process of the star-forming and evolution of massive stars.}

\newcommand{\sun}{_{\odot}}

\begin{document}
	
	\maketitle

	\section{Introduction}
	
	Almost seventy years after the pioneering work published by Salpeter \cite{salpeter}, significant progress was made both observationally and theoretically allowing a more accurate description of the initial mass function (IMF) from the hydrogen-burning limit ($M_\star\sim0.08M\sun$) all the way to the most massive stars ($M_\star\sim100M\sun$). IMF is a paramount ingredient for population synthesis models and therefore our understanding of stellar clusters and galaxy properties. 
	
	However, there are still several open theoretical questions, among them, one pointed by \cite{zinnecker} at the meeting that celebrated the 50th anniversary of the IMF and Ed Salpeter's 80th birthday, in Siena, Italy. This question was entitled: ``\textit{Is the IMF an ensemble average, due to the central limit theorem (thus log-normal)?}''. As reported by \cite{zinnecker}, this challenge, as well as others, points to the question whether the IMF is by and large universal. According to this author, it is to need more convincing IMF observations and a better theoretical understanding of why the IMF should be so robust. But what is the wider meaning of word ``robust''? Recently, \cite{cw2012} proposed a new IMF description through the stable distributions (e.g.: Gaussian distribution). As quoted by these authors and also mentioned by \cite{maschberger}, this kind of distribution considers which the star formation is a purely additive stochastic process.
	
	In fact, the IMF scenario is dominated by power-laws. See, for instance, the descriptions proposed by \cite{salpeter,scalo,kroupa1993,zannetti}. In particular, \cite{zannetti} analyzed several distributions that can adjust to IMF from lognormal to left truncated beta distributions, as well as the Pareto distribution. In addition, the distributions studied by authors are modeled using between 4 and 9 parameters and, however, affecting the goodness-of-fit tests.
	
	Such a law presents an asymptotic behavior and is valid in a specific mass range. In general, power-laws are usually found in systems that exhibit (multi)fractal behavior, long-range interactions and long-term memory that can be classified as \textit{Complex Systems}. In contrast, systems that exhibit weak interactions and no memory are better defined as a simple system. In particular, this type of system is derived from a exponential law, as it is the case of the $\log$-normal distribution. For simple systems, we applied the well-known Boltzmann-Gibbs (BG) statistical mechanics that is defined in terms of probability distribution with exponentials which emerges from Central Limit Theorem (CLT). Last two decades, there has been an increasing interest in $q$-generalized statistical mechanics\footnote{For a complete and updated list of references, see http://tsallis.cat.cbpf.br/biblio.htm.} applied successfully in different astrophysical problems, which has been achieved by researchers such as \cite{burlaga2004} and \cite{Soares06,Soares11} and, more recently, by \cite{defreitas2012,defreitas2013} showing that this framework is consistent with the dynamics that controls solar magnetic activity.
	
	Inspired by multifractal systems, \cite{tsallis1988} proposed the nonextensive entropy $S_{q}$, defined by (for further details: \cite{Abe01})
	\begin{equation}
	\label{tsallis1}
	S_{q}=k\frac{1-\int [p(x)]^{q}dx}{q-1} \quad\ (q\in \texttt{R}),
	\end{equation}
	where $q$, denoted as entropic index, is related to the degree of nonextensivity. What is $p(x)$? In this case, the entropy cannot be represented by an additive process, but a non-additive ones denoted by $S_{q}(A+B)=S_{q}(A)+S_{q}(B)+(1-q)S_{q}(A)S_{q}(B)$. We can recover the BG entropic additivity if $q=1$. As cited by \cite{johal}, the value of this index is a typical characteristic of the system, or a class of universality of the system. A strong property inherent to this generalization is, of course, to assume the Boltzmann exponential as a power-law given by
	\begin{equation} \label{tsallis2}
	\exp_{q}(x)=\left[1+(1-q)x\right]_{+}^{\frac{1}{1-q}}
	\end{equation}
	where sign ``+'' denotes $1+(1-q)x\geq0$ and 0 otherwise. For $q\rightarrow 1$, the Boltzmann exponential is recovered. This generalized exponential also known as $q$-exponential has allowed the generalization of several distribution such as the Gaussian ($q=1$) and Cauchy-Lorentz ($q=2$) distributions, as well as a generalized version of Weibull distribution obtain by \cite{picoli2003}. As mentioned by these authors, a good part of the success of $q$-distribution is due to its ability of exhibit fat-tails and power-law like behavior. Among the several $q$-distributions we focus on a special look at $q$-Weibull Distributions. The possibility to use $q$-Weibull to describe the IMF was not employed in this environment, nor any other $q$-distribution. 
	
	The main aim of the present letter is to apply $q$-Weibull distribution and to analyze some properties and details that are relevant to IMF and were not highlighted previously. Our paper is organized as follows:
	in the next section, we provide a detailed mathematical description of $q$-distribution used in our nonextensive model, the $q$-Weibull. In following, we investigate the physical implications using two well-studied clusters, and last section is the summary. 
	
	\begin{figure*}
		\begin{center}
			\includegraphics[width=0.5\textwidth, angle=90]{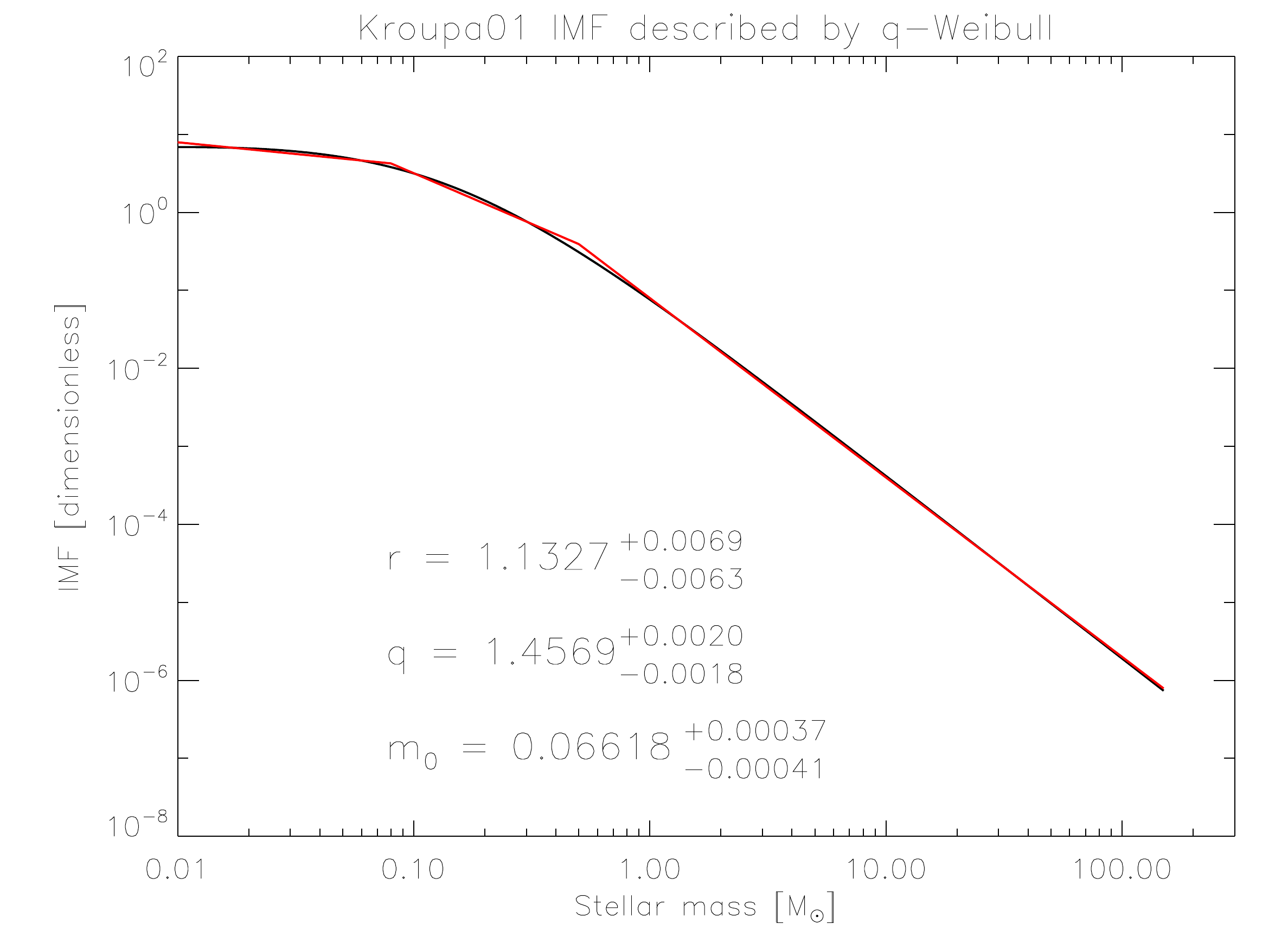}
		\end{center}
		\caption{Log-log plot of mass distribution for \cite{kroupa2001} IMF. The red line represents the best-fit using a $q$-Weibull distribution.}
		\label{fig1}
	\end{figure*}
	
	\section{Nonextensive canonical IMF}
	
	The available constraints can be conveniently summarized by the
	multiple-part canonical IMF \cite{kroupa2001} can be defined as
	\begin{eqnarray}
	\label{sal1}
	p(m)\propto m^{-\alpha_{i}},
	\end{eqnarray}
	where $\alpha_{0}=0.3\pm0.7$ if $0.01\leq m/M_{\odot}<0.08$, $\alpha_{1}=1.3\pm0.5$ if $0.08\leq m/M_{\odot}<0.5$, $\alpha_{2}=2.3\pm0.3$ if $0.5\leq m/M_{\odot}<1.00$ and while $\alpha_{3}=2.3\pm0.7$ if m/$M_{\odot}\geq1.00$ (see \cite{kroupa2001}). These values are plotted in fig.\ref{fig1} (in red).
	Specifically, for mass range $0.4\leq m/M_{\odot}<10$ the slope $\alpha=2.35$ given us the power-law distribution, widely known as Salpeter IMF. 
	
	In nonextensive scenario, probability distributions or also called $q$-distributions present important properties that can be used in context of the IMF. To represent these laws in only one distribution function, we need a stretched distribution. In the nonextensive scenario, there is a distribution function denoted by $q$-Weibull that present this behavior, where $q$ denotes the entropic index in statistical mechanics. 
	
	In this paper, we revisit the canonical IMF proposed by \cite{kroupa2001} and \cite{maschberger}. Our proposal is to use an asymmetric $q$-distribution which emerges within the nonextensive framework. We adopt the nonextensive $q$-Weibull distributions proposed by \cite{picoli2003} and recently formulated by \cite{assis2012}. The $q$-Weibull distributions are given by the Probability Density Function (PDF)
	\begin{equation}
	\label{qdis1}
	p_{q}(m)=p_{0}\frac{r}{m_{0}}\left(\frac{m}{m_{0}}\right)^{r-1}\exp_{q}\left[-\left(\frac{m}{m_{0}}\right)^{r}\right],
	\end{equation}
	for $1+(q-1)(m/m_{0})^{r}\geq 0$ and $p_{q}(m)=0$ otherwise, where $m_{0}$ is a scale parameter and $r$ a shape parameter. Note that for $r=1$ and $q\neq1$ we obtain the $q$-exponential. Respectively, when $q\rightarrow1$ with $r\neq1$ or $r=1$ we recover the Weilbull or the exponential distributions. 
	
	In general, nonextensive distributions are widely applied in systems that present long-range correlations, multifractality and asymptotic power law behavior. In particular, $q$-Weibull PDF can be used to represents skewed short and high tailed distributions, a typical feature of complex systems such as those found in astrophysical ones. The standard normalization condition for a probability is given by
	\begin{equation}
	\label{qdis1a}
	\int^{m_{u}}_{m_{l}}p_{q}(m)dm=1,
	\end{equation}
	where $m_{u}$ and $m_{l}$ are the upper and lower mass limit, respectively. According to this condition, we find that
	\begin{equation}
	\label{qdis1b}
	p_{0}=\frac{2-q}{C(m_{u})-C(m_{l})},
	\end{equation}
	where $C(x/x_{0})\equiv \left[\exp_{q}(x/x_{0})\right]^{2-q}$.
	
	Basically, Salpeter IMF is an asymptotic representation of $q$-Weibull distribution for high mass (originally Salpeter described stars up to $10\,M_{\odot}$). Furthermore, for $q>1$ and $m\gg m_{0}/(q-1)^{1/r}$, eq. (\ref{qdis1}) exhibits asymptotic behavior for different mass range. More specifically,
	\begin{equation}
	\label{qdis2}
	p_{q}(m)\sim m^{r-1},
	\end{equation}
	for low mass ($m\ll m_0$). However, this lower mass limit does not relevant to this paper, because our interesting is in asymptotic region where we have observational data. In this case, this asymptotic regime is given by
	\begin{equation}
	\label{qdis3}
	p_{q}(m)\sim m^{-r\left(\frac{2-q}{q-1}\right)+1},
	\end{equation}
	for high mass ($m\gg m_0$). In this context, from eq. (\ref{qdis3}), we can immediately correlate the high mass ($hm$) index $\alpha_{hm}$ and the nonextensive index $q$ by relationship
	\begin{equation}
	\label{qdis4}
	\alpha_{hm}=r\left(\frac{2-q}{q-1}\right)+1,
	\end{equation}
	where for $q\rightarrow 1$, we have $\alpha_{hm}\rightarrow \infty$. The low mass ($lm$) index is only associated to parameter $r$, in such a way that
	\begin{equation}
	\label{qdis5}
	\alpha_{lm}=r-1.
	\end{equation}
	
	We used a Bayesian approach for estimating the best values for the parameters $q$ and $r$. As a result, the values of $q$ and $r$ obtain in Fig.\ref{fig1} given us a $\alpha_{hm}=2.33\pm0.01$, i.e., approximately equal to Salpeter slope. Already, for low mass, we obtain $\alpha_{lm}=0.13\pm0.01$ a value slightly below the expected value of 0.3.
	
A priori, the $q$-Weibull distribution provides better adjustments than the usual Weibull one because of the addition of a new parameter $q$. However, when the parameter $q$ is close to one the $q$-Weibull distribution does not give an expressive difference when compared with that obtained from usual distribution. On the other hand, when the distribution shows heavy-tails ($q>1$), such limit is not a good approximation and, therefore, the $q$-Weibull distribution gives a significative improvement in the adjustment. In addition, the new pathway parameter $q$ facilitates a slow transition to the Weibull as $q\rightarrow 1$. Besides, the $q$-Weibull is necessary because we cannot recover the IMF’s exponents (e.g., Salpeter and Kroupa exponents) using the Weibull distribution as can be seen in Equation (\ref{qdis4}).
	
	\section{Physical implications on the index $q$}
	
	\begin{figure*}
		\begin{center}
			\includegraphics[width=0.6\textwidth]{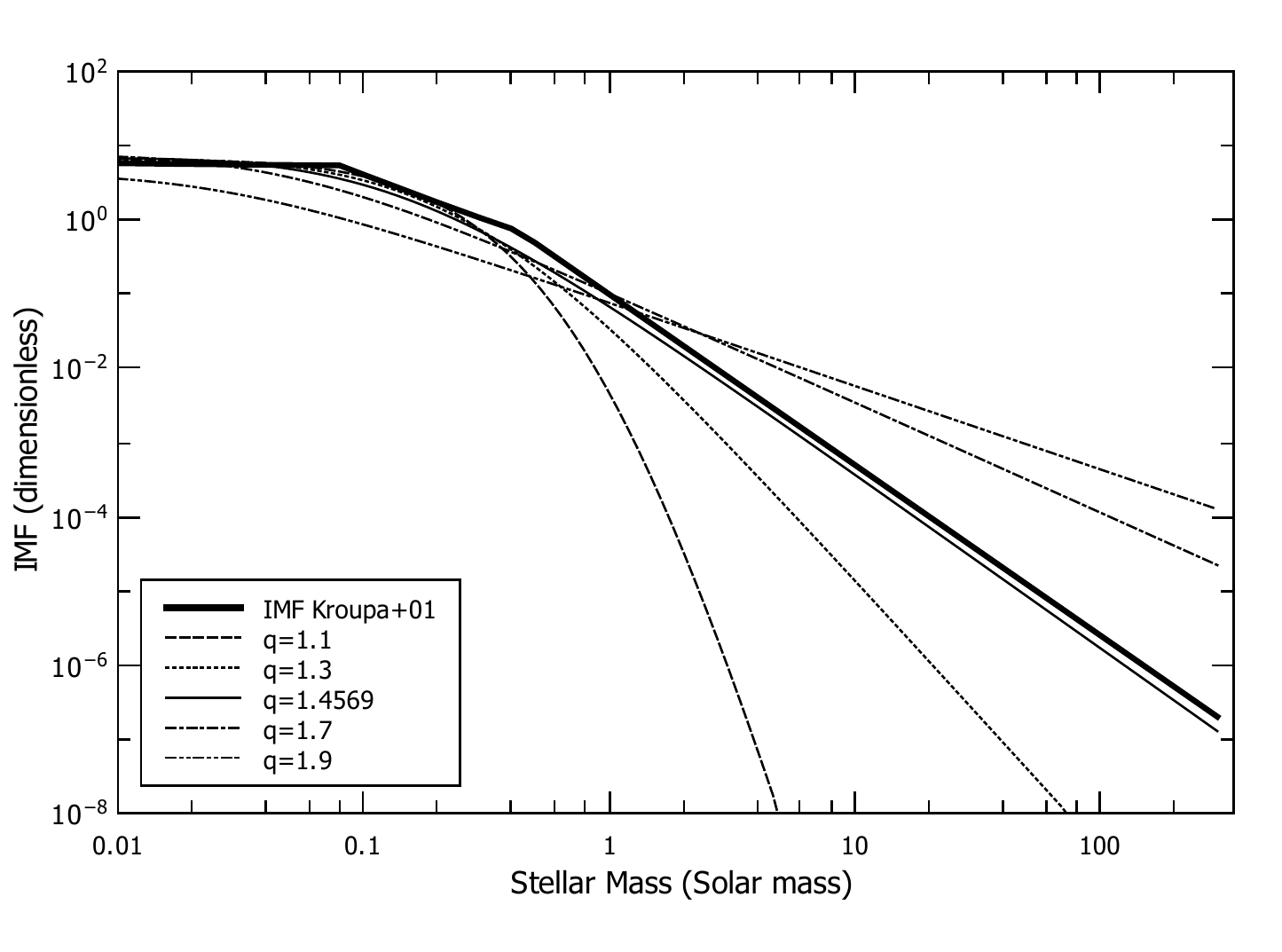}
		\end{center}
		\caption{Entropic index $q$ influence on the behavior of $q$-Weibull with parameters $r$ and $m_0$ fixed and equal to those obtained in the best-fit of figure \ref{fig1}. The mass distribution for the Kroupa IMF \cite{kroupa2001} is represented by the solid, thick line.}
		\label{fig2}
	\end{figure*}

	The use of $q$-Weibull distribution function, besides presenting a way of describing the various IMF regimes along the mass bands, has parameters that can adapt to mass functions of systems that present different behavior than the one proposed by \cite{kroupa2001}. The value of $q = 1.4569$, obtained by the adjustment of $q$-Weibull and shown in figure \ref{fig1}, points to a non-extensivity behavior in the star formation process.
	
	In figure \ref{fig2} we have the variation of the index $q$ in the interval between $1$ and $2$ with step of $0.1$ and values of $m_0$ and $r$ fixed and equal to those obtained in the best fit shown in figure \ref{fig1}. Also, according to figure \ref{fig2}, the influence caused by the variation in the values of $q$ for the region of low mass is minimal. That is, it is not this shape parameter, the one responsible for determining the amount of low mass stars that will be born in a certain region, leaving for the $r$ parameter this characteristic. In this context, it is important to remember that the IMF does not determine the number of low-mass stars. On the other hand, in the high mass region, the strong dependence of the $q$-Weibull distribution function on the variation in the values of the entropic index $q$ is clearly evident.
	
	As the behavior of the IMF in the region of massive stars is determined by the exponents $\alpha_2$ and $\alpha_3$ (see previous Section), it is expected that there is an anti-correlation between the value of these exponents and the entropic index $q$, since a growth in the value of $\alpha$ accentuates the fall in the number of high mass stars which would lead $q$ to unity. This is an interesting result as with the passing of generations of stars it is expected that the amount of high mass stars will decrease as opposed to the number of low mass stars that increase. Statistically, the system would tend to steady-state equilibrium $(q = 1)$ with the passing of the generations, which would decrease the number of massive stars and, as seen in figure \ref{fig2}, would lead to a decrease in the $q$ value, making this parameter tend to one.
	
	Generally speaking, when the system ages, more massive stars become neutron stars or black holes. In this case, there is a reduction in the number of massive stars in the distribution. As a consequence, the tail of the distribution is reduced which leads the value of $q$ to one, as shown in the curves present in Figure \ref{fig2}. However, the IMF is a probability distribution that takes into account the stars that enter the main sequence, i.e., when the hydrogen ignition is initiated. In this context, the value of computed $q$ must be greater than one, but if we consider the evolution and death of stars that occurs at different rates for different masses, $q$ is a function of time and therefore the future and present-day mass function will affect the behavior of $q$, as shown in Figure \ref{fig2}.
	
	As a way to help such discussion, let us take the alpha-plots for clusters populations from Kroupa \cite{kroupa2001}, where it is observed that the scattering of the $\alpha$ values become smaller with the increase in the number of bodies in the system, mainly for $\alpha_2$ and $\alpha_3$. It is also observed a shift to the left and slightly above the simulated values for $\alpha_2$ and $\alpha_3$ for 70\,Myr older clusters when compared to the time at which the IMF was calculated. This behavior may indicate a decrease in the effective number of massive stars as generations pass. Considering the relationship between the $\alpha_2$ and $\alpha_3$ values with the entropic index $q$, this last one can be interpreted as a regulator of the dynamic stellar evolution of massive stars and possibly must be related to age of the star-forming region, or at least as an indication of the number of generations of that region.
	
	\section{Final remarks}
	
	In our paper, we have investigated the $q$-Weibull distributions which emerge within nonextensive statistical mechanics. We verified that it has an upper limit strongly connected with the broadness stellar mass, as defined by Equation~(\ref{qdis4}). According to Figure~\ref{fig2}, higher values of $q$ are associated to heavy-tails in IMF, whereas lower values of $q$ towards to unity are linked to low-mass distributions. There is also a lower limit of the $q$-Weibull distribution associated to parameter $r$, as can be seen in Equation~\ref{qdis5}. Besides, our results also point out that the $q$-index can be interpreted as possible indicator of star-forming region age.

\acknowledgments
DBdeF acknowledges financial support 
from the Brazilian agency CNPq-PQ2 (grant No. 311578/2018-7). Research activities of STELLAR TEAM of Federal University of Cear\'a are supported by continuous grants from the Brazilian agency CNPq.

\end{document}